\documentclass[11pt,a4paper]{article}
\usepackage{graphics,graphicx}
\usepackage{amssymb,epsfig,amsmath,euscript,array,bbm}
\usepackage{axodraw}
\usepackage{cite}
\usepackage{color}

\def\beqa{\begin{eqnarray}}
\def\eeqa{\end{eqnarray}}
\def\beq{\begin{equation}}
\def\eeq{\end{equation}}

\def\one{\mbox{1 \kern-.59em {\rm l}}}
 
 \def\cA{{\cal A}}

\def\cM{{\cal M}} \def\cN{{\cal N}} 
\def\cP{{\cal P}}  \def\cR{{\cal R}}

\def\eps{{\epsilon}}

\newcounter{multieqs}

\newenvironment{pretty}{}{}

\begin{document}

\thispagestyle{empty} \setcounter{page}{0}

\begin{flushright}
QMUL-PH-09-13
\end{flushright}

\vspace{20pt}

\begin{center}

{\Large \bf A Note on Loop Amplitudes in QED}

\vspace{33pt} {\bf  Andreas Brandhuber, Gabriele Travaglini and Massimiliano Vincon}%
\begin{pretty}\footnote{{\sffamily \{\tt a.brandhuber, g.travaglini, m.vincon\}@qmul.ac.uk }}\end{pretty}

\vspace{2cm}

{\em Centre for Research in String Theory\\ Department of Physics\\ Queen Mary, University of London\\ Mile End Road, London, E1 4NS\\ United Kingdom
}\\ 

\vspace{45pt} {\bf Abstract}

\end{center}

\noindent 
We consider the two-loop four-point amplitude in  ${\cal N}=2 $ super QED, 
and show that there exists an approximate recursive structure similar to that captured 
by the ABDK/BDS ansatz for MHV amplitudes in ${\cal N}=4$ super Yang-Mills. 
Furthermore, we present a simple relation between the box coefficients of 
one-loop photon MHV amplitudes in (super) QED, 
and sums of box coefficients of one-loop MHV amplitudes 
in (super) Yang-Mills. 
\newpage


\section{Introduction} \index{1}
\setcounter{footnote}{0} 

One of the important realisations of the past decades is that physical observables in quantum gauge theories are far simpler than one would expect from  Feynman diagrams. 
For instance, the  Parke-Taylor formula \cite{Parke:1986gb} for the maximally helicity violating (MHV) scattering amplitudes in colour-ordered Yang-Mills theory at tree level resums large numbers of Feynman diagrams into a stunning one-line expression. Such intriguing  simplicity persists at the quantum level, culminating perhaps in the higher-loop iterative structures discovered in the loop expansion of maximally supersymmetric Yang-Mills (SYM) in \cite{abdk,bds}. 

The perturbative expansion of supergravity theories is also full of surprises. At tree level, there are interesting relations between amplitudes in Yang-Mills and in gravity, starting with the KLT relations \cite{Kawai:1985xq} and continuing with the recent solution of the BCF recursion relations \cite{bcfrec,bcfw} for general relativity  \cite{bbst,cs} found in \cite{Drummond:2009ge}, which expresses amplitudes in maximal supergravity in terms of sums of squares of $\cN=4$ SYM amplitudes. 
Both KLT formulae and the relations of  \cite{Drummond:2009ge} have echoes  in the expressions for the 
one-loop box coefficients \cite{Bern:2005bb,BjerrumBohr:2005xx,hall,kst}. 
Most importantly, there is now mounting  evidence of the remarkable similarities  between
$\cN=4$ SYM and $\cN=8$ supergravity, leading to  the conjecture that the
$\cN=8$ theory could be ultraviolet finite, which is supported by  multi-loop perturbative calculations
\cite{bsgf1,bsgf2,bsgf3,Bern:2009kd}.

In describing  the remarkable web of regularities and similarities between the perturbative expansions of gauge theory and gravity,  Quantum Electrodynamics (QED) has its own place in the story. 
For example,  multi-photon amplitudes in QED (with at least eight photons)  have in common  
with maximally supersymmetric  Yang-Mills and supergravity the no-triangle (and no-bubble) property. This is the 
statement that all one-loop amplitudes  can be written as sums of box functions
times rational coefficients%
\footnote{One-loop photon amplitudes in (S)QED are somewhat special since they are both infrared  and ultraviolet finite. This implies particular relations between the box coefficients, since the infrared divergences must cancel.}. This property was proven for 
$\cN=4$ SYM in \cite{bddk}, 
conjectured for $\cN=8$ supergravity in \cite{Bern:1998sv, Bern:2005bb,BjerrumBohr:2005xx, BjerrumBohr:2006yw} and subsequently proved in  \cite{BjerrumBohr:2008vc,cahk}. 
Recently it was found  in \cite{Badger:2008rn} that a similar statement holds for photon amplitudes in QED.
We also mention the  interesting connections found in \cite{bsgf3} and
\cite{BjerrumBohr:2008vc,Badger:2008rn} between the unexpected cancellations in 
one-loop scattering amplitudes,  
and the  large-$z$ behaviour of tree amplitudes observed in \cite{bbst,cs,ben,ah1,cahk}.
In unordered theories such as gravity and QED these cancellations are amplified by the
summation over different orderings of the external particles.

Two more interesting facts are worth mentioning. 
Firstly, the one-loop MHV and four-point two-loop photon amplitudes in $\cN=2$ SQED have a uniform degree of transcendentality, {\it i.e.} at one and two loops, only terms with total polylogarithmic weight equal to 2 and 4 appear, respectively \cite{Binoth:2002xg}. A similar fact has been recently observed in \cite{Naculich:2008ew,Brandhuber:2008tf} in the one- and two-loop graviton MHV amplitudes in maximal supergravity.    
Furthermore, the $\cN=2$ SQED result for these amplitudes can be obtained from  the corresponding $\cN=1$ SQED result by keeping only terms with maximal transcendentality (and no ratio of kinematical scales), leading to the speculation that maximal transcendentality \cite{klov} 
could be a feature of {\it all} maximally supersymmetric theories. 
Moreover, slightly departing from the realm of scattering amplitudes, we would also like to recall 
the somewhat puzzling ``simplicity" of the three-loop electron anomalous dimension \cite{kinoshitacvitanovic}. 
Here, numerically large cancellations occur between different diagrams, a fact which is due to the breaking of gauge symmetry at the diagrammatic level, see \cite{cvitanovic} for a prescient and  enjoyable discussion of this point.

It is therefore natural to ask to which extent the simplicity found in the perturbative expansion of amplitudes in supersymmetric Yang-Mills and supergravity persists in (S)QED.
We are fortunate to have a large number of analytic amplitudes at our disposal to test this. 
The one-loop four-photon amplitudes for massless and massive fermions were first computed in 
\cite{Karplus:1950zz,Karplus:1950zza}. Corrections to light-by-light scattering at two loops were determined 
about fifty  years later in \cite{Bern:2001dg} using the modern unitarity method \cite{bddk,fusing}. 
The four-point results of \cite{Bern:2001dg} were confirmed in \cite{Binoth:2002xg} and extended to ${\cal N}=1$ and ${\cal N}=2$ SQED by analysing the tensorial structure of the amplitudes found in \cite{Karplus:1950zz,Karplus:1950zza}. In \cite{Mahlon:1993fe}, analytic expressions for one-loop MHV photon amplitudes for an arbitrary number of photons 
were calculated with the help of the off-shell currents found in \cite{Mahlon2}. 
In \cite{Binoth:2007ca}, analytical results for all six-photon QED amplitudes were given whilst in
\cite{Bernicot:2007hs}, formulae for $n$-point MHV amplitudes in  QED, scalar QED and ${\cal N}=1$ SQED  
were obtained, together with the analytical results for the six-point NMHV QED and ${\cal N}=1$ SQED amplitudes, 
which confirmed earlier work of \cite{Mahlon:1993fe,Binoth:2007ca}.

At tree level, the simplest nonvanishing scattering amplitude one encounters in massless QED 
is the MHV amplitude with $n$ photons and two fermions,%
\footnote{There is no tree-level photon amplitude corresponding to the gluon MHV amplitude in Yang-Mills.}
\beqa
\label{mhvqed}
\cA_{\rm MHV} ( \bar{q} , q, 1^{+}, 2^{+}, \ldots , i^{-} , \ldots , n^+ ) & = & 
 i \, 
{
\langle q \,i \rangle^3 \langle \bar{q}\, i \rangle \over \langle  \bar{q}\,q \rangle^2 } 
\prod_{l=1}^n {\langle  \bar{q} \,q \rangle \over  \langle q\, l \rangle \langle l\, \bar{q} \rangle }
 \\ \nonumber  &=& 
 i \, 
{
\langle q \,i \rangle^3 \langle \bar{q}\, i \rangle \over \langle  \bar{q}\, q\rangle^2 } \sum_{\mathcal{P} \{1, 2, \ldots , n \}} 
{ \langle \bar{q}\, q\rangle \over  \langle  q\,1 \rangle \langle 1\,2 \rangle \cdots \langle n\, \bar{q} \rangle} 
\ , 
\eeqa
where  the fermion $q$ and the $i^{\rm th}$ photon have  negative helicity, and all the other particles have positive helicity. 
Equation \eqref{mhvqed}   shows two important features. Firstly, the MHV amplitude in QED is given by a compact, one-line expression, see the first line of \eqref{mhvqed}. 
Furthermore, this amplitude can be derived by summing over permutations of colour-ordered amplitudes in Yang-Mills where the photons are replaced by gluons with the same helicities.%
\footnote{See  \cite{Mangano:1990by} for a discussion of this important feature.}
This is explicitly shown in the second line of 
\eqref{mhvqed}, where each term in the sum over permutations $\cP\{1, 2, \ldots , n\}$  
is equal to a colour-ordered  Yang-Mills MHV amplitude with $n$ gluons and two fundamental fermions $q$ and $\bar{q}$. 

This observation leads directly to the first result we present in this paper.  We will discuss how the one-loop MHV amplitude of photons in supersymmetric and in pure QED can be derived directly by summing over appropriate permutations of the corresponding result for gluon MHV amplitudes in supersymmetric or pure Yang-Mills theory. As we mentioned before, one-loop photon amplitudes in (S)QED can be written in terms of (the finite parts of) box functions for $n \geq 8$. We will therefore show that the box coefficients of the Yang-Mills amplitudes, summed over appropriate permutations 
of the external gluons, directly give the box coefficients of the QED amplitudes. We will also  outline the proof of this interesting fact, based on a MHV diagram calculation \cite{csw,ozeren,bst,Bedford:2004py,Bedford:2004nh}.

The second observation we make in this paper is aimed at uncovering possible cross-order relations in the perturbative expansion of $\cN=2$ SQED.  The first example of iterative structures was 
found in planar ${\cal N}=4$ SYM for the four-point MHV amplitudes in \cite{abdk}. 
In a subsequent paper \cite{bds}, Bern, Dixon and Smirnov (BDS) put forward a conjecture for the all-loop resummation of the planar $n$-point MHV amplitudes in ${\cal N}=4$ SYM, which has been tested up to three loops in the four-point case \cite{bds} and up to two loops in the five-point case \cite{Cachazo:2006tj, Bern:2006vw}.
However, in \cite{Alday:2007he} it was realised that the BDS ansatz is incomplete at least for a large number of external gluons. 
Specifically, a direct calculation of the two-loop six-gluon MHV amplitude in 
${\cal N}=4$ SYM performed in \cite{seven} showed that the BDS ansatz breaks down, and has to be amended 
by adding a dual conformal invariant remainder function \cite{dhksbum,dhks6}. 

Motivated by this, we will consider the four-photon MHV amplitude at one and two loops in
maximally supersymmetric $\cN = 2$ SQED, 
and test the possibility that the two-loop amplitude could be 
written as a polynomial in the one-loop amplitude. 
One important difference compared to
Yang-Mills is that in QED the one- and two-loop four-photon
amplitudes are finite. Thus, one lacks the guiding principle of 
the exponentiation of infrared divergences, which is central to the all-loop ABDK/BDS ansatz. 
Despite this, we find that, quite surprisingly, the real part of the two-loop four-photon MHV amplitude in maximally supersymmetric QED is not exactly given but well approximated (in a wide kinematic region) by a polynomial in the one-loop MHV four-photon amplitude. We also discuss the limitations of such an approximate formula.

The rest of the paper is organised as follows. In Section 2 we present  the relationship mentioned earlier between the box coefficients of one-loop MHV (S)QED amplitudes and sums of  permutations of box coefficients of the same amplitudes in (S)YM, and prove it using one-loop MHV diagrams. 
In Section 3, after reviewing salient features of the BDS ansatz, we investigate 
approximate recursive structures for MHV four-photon amplitudes in ${\cal N}\!=\!2$ SQED.

\section{One-loop photon amplitudes in massless (S)QED} \index{1}

\setcounter{equation}{0} 
In this section we wish to comment on a simple relation between massless scalar QED and pure Yang-Mills amplitudes, as well as a similar one between ${\cal N}=1$ SQED and ${\cal N}=1$ SYM amplitudes. Similar relations, based on certain permutation sums of gluon amplitudes, are
known for tree amplitudes \cite{Mangano:1990by} and one-loop amplitudes (see {\it e.g.}\cite{bern1,bern2}).

We start by considering the  expressions for the  scalar QED and ${\cal N}=1$ SQED photon MHV amplitudes at one loop. These amplitudes were  computed in \cite{Bernicot:2007hs}, and 
are given by
\begin{eqnarray} 
\label{Eq1}
\mathcal{A}_{n}^{\rm scalar/{\cal N}=1} \!\!&& \!\!\!\!\!\!\!(1^{-}, \,2^{ -}, \,3^{+},\cdots ,n^{+}) \,  =\, i \frac{(e \sqrt{2})^{n}}{16 \pi^{2}}\sum_{{\cal P}\{1,2\}}
\sum_{{\cal P} \{3,\ldots, n\}}
\frac{d^{\rm scalar/{\cal N}=1}}{(n-4)!}B^{1m}(s_{23},s_{24},s_{15 \cdots n})\\ &&\!\!\!\!\!\!\!\!\!\!\!\!\!\!\!\!\!\!
+\ i \frac{(e \sqrt{2})^{n}}{16 \pi^{2}}  \sum_{{\cal P} \{1,2\}}\sum_{{\cal P}\{3,\ldots,
n\}}\sum_{m=5}^{n-1}\frac{(-1)^{m}d^{\rm scalar/{\cal N}=1}} {(n-m)!(m-4)!}B^{2me}
(s_{135 \cdots m},s_{145 \cdots m},s_{15 \cdots m},s_{2 \,  m+1 \cdots n})\, ,\nonumber
\end{eqnarray}
where
\begin{eqnarray} 
\label{coefficientsN=1}
d^{\rm scalar}&=&-2\frac{\langle 1 \, 3 \rangle \langle 1 \, 4
\rangle \langle 2 \, 3 \rangle \langle 2 \, 4 \rangle }
{ \langle 3 \,4 \rangle^{2}}\prod_{\substack{i=5 \\ i\neq 3,4}}^{n}\frac{\langle 3 \, 4 \rangle^{n-4}}
{\langle 3 \, i\rangle \langle 4 \, i \rangle} \, , 
\\ 
d^{{\cal N}=1}&=&-\langle 1 \, 2 \rangle^{2}\prod_{\substack{i=5 \\ i\neq 3,4}}^{n}\frac{\langle 3 \, 4
\rangle^{n-4}}{\langle 3 \, i\rangle \langle 4 \, i \rangle} \, .
\label{coefficientsN=0}
\end{eqnarray}
Let us explain the notation employed in \eqref{Eq1}. 
Firstly, the sums in (\ref{Eq1}) are over permutations ${\cal P}$ of the massless states inside the curly brackets. 
Secondly,  the function $B^{2me}$ appearing in \eqref{Eq1} is the
finite part of the two-mass easy scalar box functions $F$ \cite{Bern:1993kr,bst}, 
\begin{eqnarray} \label{boxes}
F^{2me}\big(s,t,P,Q) & = &
-\frac{1}{ \epsilon^{2}}\big[(-s)^{-\epsilon}+(-t)^{-\epsilon}-(-P^{2})^{-\epsilon}-(-Q^{2})^{-\epsilon}\big]
+B^{2me}\big(s,t,P,Q\big)\nonumber \, ,
\end{eqnarray}
where
\begin{eqnarray}
B^{2me}\big(s,t,P^{2},Q^{2}\big) & = &
\textrm{Li}_{2}\big(1-aP^{2}\big)+\textrm{Li}_{2}\big(1-aQ^{2}\big)-\textrm{Li}_{2}\big(1-as\big)-\textrm{Li}_{2}\big(1-at\big)\nonumber \, ,
\end{eqnarray}
with
\begin{equation}
a\, :=\, \frac{P^{2}+Q^{2}-s-t}{P^{2}Q^{2}-st}\, .
\end{equation}
As usual,  $s:=(P+p)^{2}$, $t:=(P+q)^{2}$, with $p+q+P+Q=0$, where
$p$ and $q$ are the  massless legs (sitting at opposite corners, in the two-mass easy boxes), and $P$ and $Q$ the  massive legs.  
The arguments of the box functions appearing in \eqref{Eq1} are the kinematical invariants  $s_{i \cdots j}:=(k_{i}+\cdots+k_{j})^{2}$.

In Figure 1 we provide a representation of the box function appearing in the second line of \eqref{Eq1}. The massless legs correspond to the positive helicity photons $3^+$ and $4^+$. The negative helicity photons $1^-$ and $2^-$ are always part of (different)  massive corners $P$ and $Q$, which contain $m-3$ and $n-m+1$ legs respectively. The combinatorial coefficients appearing in \eqref{Eq1} correspond to the number of permutations of the positive helicity photons inside $P$ and $Q$ (which obviously leave the box function invariant).

\begin{figure}
\begin{center}
\begin{picture}(160,185)(40,5)

\Line(80,50)(60,30)
\Line(80,50)(70,23)
\Line(80,50)(50,60)
\Line(80,50)(166,50)
\Line(80,50)(80,136)
\Text(76,21)[tr]{ {$1^{-}$}}
\Text(58,32)[tr]{ {$5^{+}$}}
\Text(53,62)[br]{ {$m^{+}$}}

\DashCArc(80,50)(20,165,225){2}

\Line(80,96)(80,136)
\Line(80,136)(55,161)
\Line(80,136)(166,136)
\Text(65,165)[br]{ {$4^{+}$} }

\Line(166,136)(181,166)
\Line(166,136)(189,160)
\Line(166,136)(196,126)
\Text(179,168)[bl]{$2^{-}$}
\Text(191,162)[lb]{$m+1^{+}$}
\Text(198,128)[lt]{$n^{+}$}
\DashCArc(166,136)(20,340,45){2}

\Line(166,136)(166,50)

\Line(166,90)(166,50)
\Line(166,50)(191,25)
\Line(166,50)(126,50)

\Text(193,23)[tl]{$3^{+}$}
\end{picture} 
\end{center}
\caption{\em The two-mass easy box function appearing in \eqref{Eq1}. The momenta $p=k_{3}$ and $q=k_{4}$ are null, whereas $P:=k_{1}+k_{5}+\cdots+k_{m} $ and 
$Q:=k_{2}+k_{m+1}+\cdots+k_{n} $ are massive.  The one-mass box function in \eqref{Eq1} is obtained by setting $m=n$, so that the top right corner becomes massless (and contains only the momentum $k_2$). }
\end{figure}
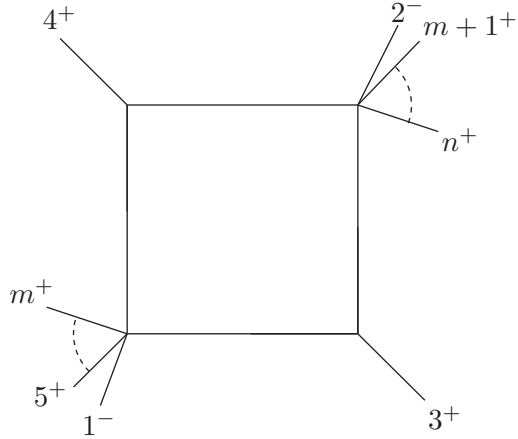

In  \eqref{coefficientsN=1} and  \eqref{coefficientsN=0},  we have multiplied the result  of \cite{Bernicot:2007hs} 
for $d^{\rm scalar}$ by a factor of $2$ to account for the fact that we are working with complex scalar fields. 
Finally, let us stress that the amplitudes given in \eqref{Eq1} are infrared and ultraviolet finite. Because of Furry's theorem, they are nonvanishing only for $n$ even. 

We now turn to the corresponding planar MHV amplitudes in Yang-Mills. 
In  ${\cal N}=4 $ SYM they were first derived by Bern, Dixon, Dunbar and Kosower in  \cite{bddk} using  unitarity \cite{fusing} and collinear limits, and later confirmed in \cite{bst} using one-loop MHV diagrams. 
They have the following form 
\begin{equation} 
\label{N4}
\mathcal{A}_{n}^{{\cal N}=4}\ = \ {\cal A}_{n}^{\rm tree} \, \sum_{p={j+1}}^{i-1} \sum_{q=i+1}^{j-1} F^{2me}(p,q,P,Q) \, , 
\end{equation}
\noindent
and   
\begin{equation}
\label{pt}
\mathcal{A}^{\rm tree}_{n}(1^{+},\ldots,i^{-},\ldots,j^{-},\ldots , n^{+}): \ = \ i \frac{\langle i \, j \rangle^{4}}{\langle 1\, 2\rangle \langle 2 \, 3\rangle \cdots \langle n \,  1\rangle}\, ,
\end{equation}
is the tree-level amplitude, given by the Parke-Taylor formula \cite{Parke:1986gb}. 

The one-loop MHV amplitude in  ${\cal N}=1$ SYM was presented in \cite{fusing} and rederived in 
\cite{Bedford:2004py} using MHV diagrams.  The contribution to the amplitude 
of an $\cN=1$ chiral multiplet running in the loop  
is given by the following compact formula,
\begin{equation} \label{N1}
\mathcal{A}_{n}^{{\cal N}=1}(1^{+},\ldots,i^{-},\ldots,j^{-},\ldots,n^{+})\, = \, 
\sum_{p=j+1}^{i-1} \sum_{q=i+1}^{j-1} [c^{{\cal N}=1}]^{ij}_{pq}\, B^{2me}\big(p,q,P,Q\big)+\cdots \, ,
\end{equation}
where  
\begin{equation} \label{c1}
[c^{{\cal N}=1}]^{ij}_{pq}\ = \  \frac{1}{2} \mathcal{A}_{n}^{\rm tree} \, b_{pq}^{ij} \, ,
\end{equation}
and
\begin{equation} \label{bij}
b^{ij}_{pq}\ = \ 2 \frac{\langle i \, p \rangle \langle j \, q \rangle \langle i \, q \rangle \langle j \, p \rangle} 
{\langle i \, j \rangle^{2} \langle p \, q \rangle^{2}}  \, .
\end{equation}
\noindent
The dots in (\ref{N1}) stand for triangle and bubble functions, which do not enter our discussion.%
\footnote{This is because of the no-triangle property of  QED amplitudes \cite{Badger:2008rn}, which ensures that 
the bubble and triangle coefficients vanish.}

Lastly, the one-loop $n$-point non-supersymmetric Yang-Mills MHV amplitudes were computed in 
\cite{fusing,Bern:1993mq,Bedford:2004nh}
and confirmed in \cite{Brandhuber:2008cy} using generalised unitarity 
\cite{Bern:1997sc,bcfgen}, with the result 
\begin{equation} \label{N0}
\mathcal{A}_{n}^{\rm scalar} (1^{+},\ldots,i^{-},\ldots,j^{-},\ldots,n^{+})\, = \,  
  \sum_{p={j+1}}^{i-1} \sum_{q=i+1}^{j-1} [c^{\rm scalar}]^{ij}_{pq} \, B^{2me}\big( p,q,P,Q \big)+\cdots\, ,
\end{equation}
where
\begin{equation} \label{c0}
[c^{\rm scalar}]^{ij}_{pq}\, =\,   \frac{1}{2} \mathcal{A}_{n}^{\rm tree} \, [b^{ij}_{pq}]^{2} \, .
\end{equation}
The dots in (\ref{N0}) stand for triangle and bubble  functions, as well as for the rational terms of the amplitude, which are not relevant in the following. 

Next we wish to expose a simple
relation between the coefficients $d^{\rm scalar/{\cal N}=1}$ of the two-mass easy box 
functions in the expression for the MHV photon scattering amplitudes 
in (S)QED, given  in  (\ref{Eq1}), 
and the corresponding coefficients of the same box function, $c^{\rm scalar/{\cal N}=1}$,  of the MHV gluon amplitudes in $\mathcal{N}=1$ and pure Yang-Mills  in (\ref{c1}) and (\ref{c0}).  
In order to match a gluon  amplitude to a target (S)QED amplitude, we introduce appropriate sums over permutations as follows.  
With reference to the box function in Figure 1, 
we hold the two massless legs fixed, and sum over permutations of the gluons appearing at the massive corners.  
The sum can be performed using the eikonal identity  \cite{Mangano:1990by}, see also  \eqref{mhvqed}, to get 
\begin{equation}
 \label{eikonalII}
\sum_{{\cal P}_{1}}\sum_{{\cal P}_{2}}
\frac{\langle 1\, 2 \rangle^{4} }{\langle 3\, 2 \rangle\langle 2 \,(m+1)\rangle \cdots \langle (n-1)\,n\rangle \langle n\, 4\rangle
\langle 4 1\rangle \langle 1 5 \rangle \cdots \langle m 3\rangle }  =\frac{\langle 1 \,2  \rangle^{4} \langle 3 \,4 \rangle^{n-4}}{\langle 3 \,1 \rangle \langle 1\,4 \rangle \cdots 
\langle 3\, n \rangle \langle n \,4 \rangle} \, ,
\end{equation}
where ${\cal P}_{1}:={\cal P}\{2,m+1,\ldots,n\}$ and ${\cal P}_{2}:={\cal P}\{1,5,\ldots,m\}$ are permutations of the massless 
legs in the massive corners of the box function. 
Multiplying (\ref{eikonalII}) by  $b^{12}_{34}$
we recover the expression for $[d^{{\cal N}=1}]^{12}_{34}$  given in (\ref{coefficientsN=1}). 
A similar argument runs for the one-mass box coefficients  $[d^{\rm scalar}]^{ij}_{pq}$, with the only difference that the sum in \eqref{eikonalII} is  over one set of permutations rather than two.

One can arrive at the same conclusion by performing a one-loop MHV diagram calculation
akin to \cite{bst,Bedford:2004py,Bedford:2004nh} with MHV rules adapted to QED as done at
tree level in \cite{ozeren}. For a one-loop MHV photon amplitude we have to glue two tree-level
MHV vertices with two internal scalar propagators, and perform an appropriate loop integration \cite{bst}.
We will not give details of the calculation because we can recycle results from \cite{bst,Bedford:2004py,Bedford:2004nh}. 
The crucial observation is that the only diagrams contributing are those
where all gluons are external and the two internal legs of each MHV vertex are either scalars or
fermions with opposite helicity. This also implies that the two external negative helicity gluons 
must belong to  different MHV vertices. The relevant MHV QED tree amplitudes can be obtained from 
the corresponding MHV tree amplitude in QCD with $n-1$ positive helicity gluons, one negative helicity gluon and two fermions (scalars) of opposite helicity and summing over the $n!$ permutations of the $n$ gluons. 
Writing the QED MHV vertices with two fermions and $n$ gluons in terms of QCD tree MHV vertices, see the second line of   \eqref{mhvqed}, 
reduces the calculation to sums of permutations of MHV one-loop diagrams for MHV amplitudes in
pure Yang-Mills and $\mathcal{N}=1$ SYM \cite{Bedford:2004py,Bedford:2004nh}. It can be easily seen that this reproduces exactly the observations made earlier in this section on the box coefficients of the one-loop (S)QED MHV amplitudes. Triangle and box coefficients are guaranteed to vanish because of the no-triangle property \cite{Badger:2008rn}.

Finally, we observe that this has implications for the twistor-space localisation properties of the coefficients, which are inherited from those of the (S)YM amplitudes,
{\it i.e.} the coefficients localise on sets of two, possibly intersecting lines in twistor space. 
It would be interesting to see whether similar structures appear in non-MHV amplitudes and at higher loops.


\section{Approximate iterative structures  in ${\cal N}=2$ SQED} \index{1}
\setcounter{equation}{0}
Motivated  by the existence of iterative structures for amplitudes in ${\cal N}=4$ SYM, 
we have investigated the possible existence of recursive-like structures for MHV amplitudes in 
the maximally supersymmetric ${\cal N}=2$ SQED theory.%
\footnote{A similar analysis has been performed in 
\cite{Naculich:2008ew,Brandhuber:2008tf}
for the four-point MHV amplitude in $\mathcal{N}=8$ supergravity, and highlighted a remarkably simple structure for the two-loop term in the expansion of logarithm 
of the helicity-blind ratio $\mathcal{M}^{\mathcal{N}=8} / \mathcal{M}_{\rm tree}$. 
The functions appearing in the ratio were also found to have uniform transcendentality.
}
Before discussing our results, let us briefly review the iterative relations in $\cN=4$ SYM \cite{abdk,bds}. 
It was shown in \cite{abdk} that  the two-loop four-point MHV amplitude in ${\cal N}=4$ SYM satisfies an intriguing cross-order relation, 
\begin{equation} 
\label{abdk}
{\cal M}^{(2)}_{4}(\epsilon)-\frac{1}{2}\big( {\cal M}^{(1)}_{4}(\epsilon) \big)^{2}=f^{(2)}(\epsilon){\cal M}^{(1)}_{4}(2 \epsilon)
+C^{(2)}+{\cal O}(\epsilon)\, . 
\end{equation}
Here $\cM_n^{(L)}$ is the helicity-blind function obtained by taking the ratio between the $L$-loop MHV amplitude and the corresponding tree amplitude. Furthermore  $f^{(2)}(\epsilon)=-(\zeta_{2}+\zeta_{3} \epsilon+\zeta_{4} \epsilon^{2})$, and 
$
C^{(2)}=-(5/4)\, \zeta_{4}$.

In \cite{bds},  a resummed, exponentiated expression for the
scalar function $\cM_{n}$ was proposed, and  checked explicitly  in a three-loop calculation in
the four-point case.  The BDS conjecture is  expressed as \cite{bds}
\beq
\label{bds}
\cM_n \ := \ 1 + \sum_{L=1}^{\infty} a^L \cM_{n}^{(L)} (\epsilon )  \ =  \
\exp \Big[ \sum_{L=1}^{\infty} a^L  \Big( f^{(L)} (\epsilon) \cM_{n}^{(1)} ( L \epsilon )  + C^{(L)} + E_n^{(L)}(\epsilon )\Big)
\Big]
\ ,
\eeq
where $a=[{g^2 N/ (8 \pi^2)}] (4\pi e^{-\gamma})^\eps$ .
Here $f^{(L)}(\epsilon )$ is a set of functions,
\beq \label{fleps}
f^{(L)}(\epsilon ) \, :=\, f_0^{(L)} + f_1^{(L)} \epsilon + f_2^{(L)} \epsilon^2  \ ,
\eeq
one at each loop order, which appear in the exponentiated all-loop expression
for the  infrared divergences in generic amplitudes in dimensional regularisation 
\cite{ir6} (and generalise the function $f^{(2)}$  in \eqref{abdk}).
In particular, $f_0^{(L)} = \gamma_{K}^{(L)} / 4$, where $\gamma_{K}$ is the cusp anomalous dimension,
related to the anomalous dimension of twist-two operators at  large spin. 
Importantly,  the constants $C^{(L)}$,  $f_0^{(L)}$, $f_1^{(L)}$ and $f_2^{(L)}$
on the right hand side of \eqref{bds} do not depend either on kinematics or on the number of particles $n$.
On the other hand, the non-iterating contributions  $E_n^{(L)}$ depend explicitly on $n$, but vanish as $\epsilon \to 0$.

BDS also suggested  a resummed expression for the appropriately defined finite part of the $n$-point MHV amplitude,
\beq
\label{finite}
\mathcal{F}_n \ =  e^{F^\mathrm{BDS}_n   }
\ ,
\eeq
where
\beq
\label{fbds}
F^\mathrm{BDS}_n (a) = \, {1\over 4} \gamma_K(a)\, \, F^{(1)}_n (0)  + C(a)
\ .
\eeq
Notice that the entire dependence on kinematics of the BDS ansatz  enters through
the finite part of the one-loop box function,
$F^{(1)}_n (0)$.

In analogy with the BDS ansatz, we would like to investigate the existence of cross-order relations in the four-point amplitude 
${\cal M}_{4}(1^{-},2^{-},3^{+},4^{+})$ in ${\cal N}=2$ SQED. 
To this end, we consider a decomposition of the two-loop term in the expansion of  this  amplitude as 
\begin{equation} 
\label{ansatzQED}
[{\cal M}_{4}^{(2)}]_{\rm ansatz}= b \,\big[{{\cal M}_{4}^{(1)}}\big]^{2} + c\, {\cal M}_{4}^{(1)} + d \, ,
\end{equation}
where ${\cal M}_{4}^{(1)}$ is the four-point one-loop MHV amplitude, and  $b,c$ and $d$ have to be determined.

The  expressions for photon-photon scattering amplitudes at one and two loops  entering \eqref{ansatzQED} are taken from \cite{Binoth:2002xg}. 
The one-loop four-photon MHV amplitude in ${\cal N}=1$ SQED is given by
\begin{equation} \label{1loop}
{\cal M}_{4}^{(1)}=-4 \big[(X-Y)^{2}+\pi^{2}\big]
\, ,
\end{equation}
where
\begin{equation} \label{XY}
X=\log \Big(\frac{-t}{s}\Big)\ ,  \quad Y=\log \Big(\frac{-u}{s}\Big)\, .
\end{equation}
A few comments are in order here. 
Firstly, we notice that in the physical region $s > 0$ and $t,u<0$ the expression (\ref{1loop}) is real. 
Outside this region,  an analytic continuation is needed as the $u$- and $t$-channels develop a discontinuity. 
Secondly, we observe in (\ref{1loop}) that ratios of kinematic scales such as $t/s$ only
appear as arguments of logarithms. 
Furthermore, all the functions appearing in the expression for  
${\cal M}_{4}^{(1)}$ have 
uniform degree of transcendentality equal to  2.

The two-loop expression for the four-point MHV ${\cal N}=2$ SQED amplitude is still rather compact and simple. 
It is given by \cite{Binoth:2002xg}
\begin{eqnarray} 
\label{2loop}
{\cal M}_{4}^{(2)}& = &-16 \, \textrm{Li}_{4}(y)+8 \, Y \, \textrm{Li}_{3}(x)+8\, Y \, \textrm{Li}_{3}(y)+\frac{16}{45} \pi^{4}\\ &&{}-\frac{2}{3}\,
X\, Y\pi^{2}-\frac{2}{3}\, Y^{3}\,\big(Y-4\,X\big) \nonumber \\ 
&&{}+i \pi \Big[16 \, \textrm{Li}_{3}(x)-\frac{4}{3}\,Y
\pi^{2}-\frac{4}{3}\,Y^{2}\big(Y-3\,X\big)\Big]+\Big\{ u \leftrightarrow t \Big\}
\nonumber\, ,
\end{eqnarray}
where $X$ and $Y$ are defined in (\ref{XY}) and 
\beq
x\, :=\, -t/s \ , \qquad  y\, :=\, -u/s\, =\, 1-x 
\ . 
\eeq
As in (\ref{1loop}),  also in \eqref{2loop} there are no terms proportional to ratios of Mandelstam variables, and in \eqref{2loop} we only have functions with  transcendentality equal to 4. Therefore, we expect the coefficients $b$, $c$ and $d$ to have transcendentality $0$ and $2$ and $4$, respectively. In contrast to the one-loop amplitude the two-loop amplitude develops an imaginary part.

The real part of (\ref{2loop}) can be recast in the following suggestive form,
\begin{eqnarray} \label{continued}
\mathbbm{Re}\Big[{\cal M}_{4}^{(2)}\Big]&\!\! = \!\!&-16 \, \textrm{Li}_{4}(y)-16\,\textrm{Li}_{4}(x)+8 (X+ Y) \, \big(\textrm{Li}_{3}(x)+
\textrm{Li}_{3}(y)\big)+4 X^{2}Y^{2}-\frac{4}{3}X Y \pi^{2}{} \nonumber \\
& & {}-\frac{1}{24}\Big[{\cal M}_{4}^{(1)}\Big]^{2}-\frac{\pi^{2}}{3}{\cal M}_{4}^{(1)}+\frac{2}{45}\pi^{4} \, .
\end{eqnarray}
In \eqref{continued}, we have re-written the result of \cite{Binoth:2002xg} 
in a way that already incorporates the functions ${\cal M}_{4}^{(1)}$ and $\big[{\cal M}_{4}^{(1)}\big]^{2}$ 
appearing in  \eqref{ansatzQED},  see  the last line of (\ref{continued}). In the following we will
focus only on the real part of $ {\cal M}_{4}^{(2)} $.

In order to find a set of coefficients $b,c$ and $d$ which best fit our proposed ansatz (\ref{ansatzQED}), we build a system of three equations for some three
random values of $y$ and solve for $b,c$ and $d$. We then plug the values of these coefficients in (\ref{ansatzQED}) and match the ansatz against the 
real part
of (\ref{2loop}). In Figure 2 the real part of (\ref{2loop}) and our ansatz are plotted. 
The values of $b, c$ and $d$ used in the plot are given by the set of values $I$ for $y$ shown 
in Table 1 below,

\begin{table}[th]
\begin{center}
\begin{tabular}{|c|r|r|r|r|r|r|r|}
\hline Coefficients & $I$ & $II$ & $III$ & $IV$ & $V$ & $ VI $ & $VII$
\\
\hline  $b$   & -0.0386  & -0.038 & -0.0387  &  -0.0417 & -0.0412 & -0.0415 & -0.0412
\\
\hline  $c$   & -2.567 &   -2.571 & -2.574 & -2.812 & -2.894 & -3.009 & -2.877
\\
\hline  $d$   & -5.784  & -5.894 & -5.938 & -11.46 & -14.689 & -25.098 & -14.489
\\
\hline $F(b,c,d)^{{\cal N}=2}$ &  97.179 & 187.129  & 88.452  & 49.555  & 1.065  & 25.554  & 0.500
\\
\hline 
\end{tabular}
\caption{\it Values of $b$, $c$ and $d$ for different sets of values for $y$. 
The sets $I$-$III$ include points away from the boundary $y=0$ and $y=1$, and the resulting coefficients $b$, $c$, $d$ 
vary slowly from one set to another, unlike the case of the sets $IV$-$VI$, which include points near the boundary in $y$ space. The set $VII$ is obtained using the least square method.
 }
\end{center}
\end{table}
\noindent
where the sets $I$--$VI$ are%
\footnote{Since (\ref{2loop}) is  symmetric under $x \to 1-x$, 
we only choose values for $y$ from half of the phase space.} 
$I=\{0.3,0.4,0.5\}$, $II=\{0.272,0.342,0.482\}$, $III=\{0.25,0.35,0.45\}$, $IV=\{0.00001,0.10001,0.482\}$, 
$V=\{0.0006,0.0023,0.006\}$, $VI=\{0.0001,0.0003,0.0235\}$, 
while the values of $(b,c,d)$ in column $VII$ are obtained applying  the least square method. We have carried out a similar analysis for the four-point MHV  amplitude in ${\cal N}=1$ SQED, the expression of which can also be found in \cite{Binoth:2002xg}, and we report the results in Table 2 below,
\begin{table}[th]
\begin{center}
\begin{tabular}{|c|r|r|r|r|r|r|r|}
\hline Coefficients & $I$ & $II$ & $III$ & $IV$ & $V$ & $ VI $ & $VII$
\\
\hline  $b$   & -0.0356  & -0.0356 & -0.0356  &  -0.0277 & -0.027 & -0.026 & -0.0278
\\
\hline  $c$   & -1.829 &   -1.828 & -1.827 & -1.062 & -0.113 & 0.211 & -0.719
\\
\hline  $d$   & -79.04  & -78.99 & -78.98 & -60.99 & 18.39 & 32.54 & -44.57
\\
\hline $F(b,c,d)^{{\cal N}=1}$ &  756.0 & 749.7  & 746.5  & 99.4 & 1265.9  & 1365.6  & 24.1
\\
\hline 
\end{tabular}
\caption{\it Values of $b$, $c$ and $d$ for different sets of values for $y$. 
The sets of $y$ used are the same as in the ${\cal N}=2$ theory shown in Table 1.
 }
\end{center}
\end{table}

Introducing 
\begin{equation} \label{leastsq}
 F(b,c,d)\, :=\, 
  \int_{0}^{1}\!\!dy\, \Big({\cal M}^{(2)}_4 (y)-[{\cal M}^{(2)}_4]_{\textrm{ansatz}}(y)\Big)^{2} \, ,
\end{equation}
and minimising $F(b,c,d)$ gives the ``best" values for $b$, $c$ and $d$ over all of the phase space. 

Let us summarise the outcome of this analysis.  

{\bf 1.} 
Unlike the case of the BDS ansatz for $\cN=4$ SYM, we find that the coefficients $b,c$ and $d$ are not (kinematic-independent) constants, but have different values for different kinematic points.

{\bf 2.} 
Having derived values of these three coefficients, we plot the ${\cal N}=2$ two-loop amplitude as well as our ansatz as functions of the ratio $y$. 
A particular example is presented in Figure 2, which shows a very surprising overlap of the
two functions.

\begin{figure}
\label{figure4}
\begin{center}
\scalebox{1.2} {\includegraphics{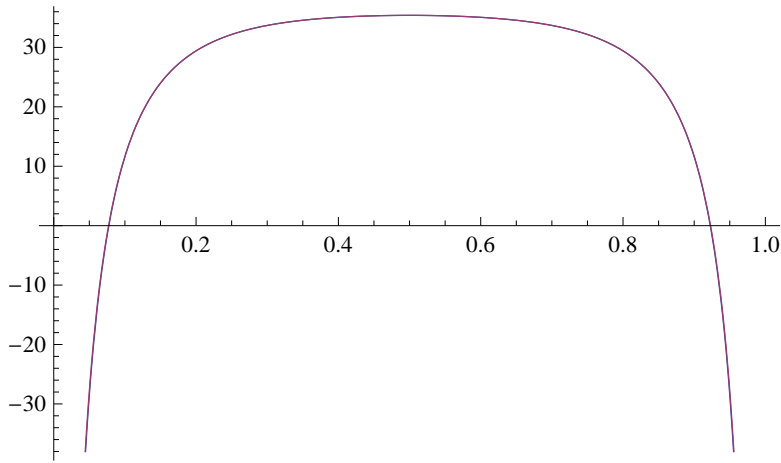}}
\end{center}
\caption{\em In this Figure we plot (the real part of) the right-hand side of \eqref{2loop}, representing the four-point two-loop MHV amplitude in  ${\cal N}=2$ SQED, together with our ansatz \eqref{ansatzQED}, with $y=-u/s$ given by set $I$.
The two overlapping functions are indistinguishable in this plot.}
\end{figure}

{\bf 3.} 
In order to study more closely the functional forms of the two-loop amplitude and of our ansatz, 
in Figure 3 we present a plot of the difference between (\ref{ansatzQED}) and (\ref{continued}), which we could 
consider as the ``remainder function" for this amplitude,%
\footnote{Notice that this function depends however on the choice of $b$, $c$ and $d$ we make in the ansatz \eqref{ansatzQED}. } 
\beq
\cR_4(y)\, := \, \mathbbm{Re}\Big[ \cM_4^{(2)} \Big] - \Big(
 b \,\big[{{\cal M}_{4}^{(1)}}\big]^{2} + c\, {\cal M}_{4}^{(1)} + d \Big)
\ . 
\eeq
From Figure 3, we see that the two functions agree for a wide range of values of $y$, 
however the difference function has spikes as $y \rightarrow 0$ or $y \rightarrow 1$. 
In these limits the best fit would be given by the second line of \eqref{continued}
in which case the disagreement would be proportional to a single power of a logarithm
in $x$ or $y$.
The presence of this divergent behaviour  at the extrema of the phase space shows that our ansatz is certainly incomplete. However, we find it quite remarkable that large deviations only appear at the boundary of the phase space. 

{\bf 4.}
We have computed (\ref{leastsq}) for both ${\cal N}=1$ and ${\cal N}=2$ SQED and found the values
\begin{equation}
F(b,c,d)^{{\cal N}=2}=0.5 \, , \quad \quad F(b,c,d)^{{\cal N}=1}=24.1\, .
\end{equation}
This shows that \eqref{ansatzQED} gives the most accurate approximation of the real part
of the two-loop amplitude in the case of ${\cal N}=2$ SQED.

{\bf 5.}
We also observe that the four-point MHV  amplitude in ${\cal N}=2$ SQED can be derived from the ${\cal N}=1$ SQED amplitude by keeping maximally trascendental terms and deleting contributions which multiply ratios of the kinematics invariants.

\begin{figure}
\label{figure6}
\begin{center}
\scalebox{1.2} {\includegraphics{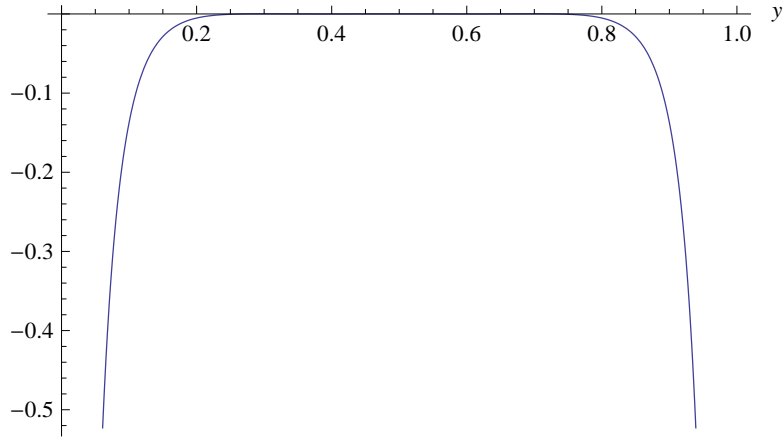}}
\end{center}
\caption{\em 
This figure shows a plot of the ${\cal N}=2$ SQED remainder function, constructed as the difference between the 
two-loop MHV amplitude in  ${\cal N}=2$ SQED and \eqref{ansatzQED} using set $I$.}
\end{figure}

In summary, the approximate iterative structures we have explored  in this section are certainly not 
on the same footing as those in ${\cal N}=4$ SYM found in \cite{abdk,bds}. Nevertheless, we find it intriguing that part of the two-loop four-photon MHV amplitude \eqref{2loop} is captured by a polynomial in the one-loop amplitude \eqref{1loop}. 
It would be interesting if one could find exact iterative structures, written in terms of an appropriate QED remainder function, and explain them in terms of some underlying symmetry of the theory. It would also be interesting to find a Wilson loop interpretation of MHV scattering amplitudes in SQED, 
similarly to that found in $\cN=4$ SYM \cite{am,dks1,bht}.

\section*{Acknowledgements}
We would like to thank Zvi Bern for helpful comments on a preliminary version of this paper
and for very useful discussions.
MV would like to thank the IPM in Tehran for their warm hospitality and 
support during the intermediate stages of this project.
This work was supported by the STFC under a Rolling Grant  ST/G000565/1.
GT is supported by an EPSRC Advanced Research Fellowship EP/C544242/1
and by an EPSRC Standard Research Grant EP/C544250/1.



\begin{thebibliography}{99}



\bibitem{Parke:1986gb}
S.~J.~Parke and T.~R.~Taylor,
{\it An Amplitude for $n$ Gluon Scattering},  Phys.\ Rev.\ Lett.\  {\bf 56}, 2459 (1986).



\bibitem{abdk}
  C.~Anastasiou, Z.~Bern, L.~J.~Dixon and D.~A.~Kosower,
  {\it Planar amplitudes in maximally supersymmetric Yang-Mills theory,}
  Phys.\ Rev.\ Lett.\  {\bf 91}, 251602 (2003),
  {\tt hep-th/0309040}.



\bibitem{bds}
  Z.~Bern, L.~J.~Dixon and V.~A.~Smirnov,
  {\it Iteration of planar amplitudes in maximally supersymmetric Yang-Mills
  theory at three loops and beyond}, Phys.\ Rev.\  D {\bf 72}, 085001 (2005),
  {\tt hep-th/0505205}.




\bibitem{Kawai:1985xq}
  H.~Kawai, D.~C.~Lewellen and S.~H.~H.~Tye,
  {\it A Relation Between Tree Amplitudes Of Closed And Open Strings,}
  Nucl.\ Phys.\  B {\bf 269} (1986) 1.



  \bibitem{bcfrec}
  R.~Britto, F.~Cachazo and B.~Feng,
 {\it New recursion relations for tree amplitudes of gluons,}
  Nucl.\ Phys.\  B {\bf 715} (2005) 499,
  {\tt hep-th/0412308}.

 \bibitem{bcfw}
  R.~Britto, F.~Cachazo, B.~Feng and E.~Witten,
  {\it Direct proof of tree-level recursion relation in Yang-Mills theory,}
  Phys.\ Rev.\ Lett.\  {\bf 94} (2005) 181602,
  {\tt hep-th/0501052}.



\bibitem{bbst}
  J.~Bedford, A.~Brandhuber, B.~Spence and G.~Travaglini,
  {\it A recursion relation for gravity amplitudes,}
  Nucl.\ Phys.\  B {\bf 721} (2005) 98,
  {\tt hep-th/0502146}.


\bibitem{cs}
  F.~Cachazo and P.~Svr\v{c}ek,
  {\it Tree level recursion relations in general relativity,}
 {\tt hep-th/0502160}.



\bibitem{Drummond:2009ge}
  J.~M.~Drummond, M.~Spradlin, A.~Volovich and C.~Wen,
  {\it Tree-Level Amplitudes in ${\cal N}=8$ Supergravity,}
  {\tt 0901.2363 [hep-th]}.





\bibitem{Bern:2005bb}
  Z.~Bern, N.~E.~J.~Bjerrum-Bohr and D.~C.~Dunbar,
  {\it Inherited twistor-space structure of gravity loop amplitudes,}
  JHEP {\bf 0505} (2005) 056,
  {\tt hep-th/0501137}.




\bibitem{BjerrumBohr:2005xx}
  N.~E.~J.~Bjerrum-Bohr, D.~C.~Dunbar and H.~Ita,
  {\it Six-point one-loop ${\cal N}=8$ supergravity NMHV amplitudes and their IR
  behaviour,}
  Phys.\ Lett.\  B {\bf 621} (2005) 183,
  {\tt hep-th/0503102}.



\bibitem{hall}
  A.~Hall,
  {\it Supersymmetric Yang-Mills and Supergravity Amplitudes at One Loop,}
  {\tt 0906.0204 [hep-th]}.
  

\bibitem{kst}
P.~Katsaroumpas, B.~Spence and G.~Travaglini,
  {\it One-loop ${\cal N}=8$ supergravity coefficients from ${\cal N}=4$ super Yang-Mills,}
  {\tt 0906.0521 [hep-th]}.


\bibitem{bsgf1}
  Z.~Bern, L.~J.~Dixon and R.~Roiban,
{\it Is ${\cal N}=8$ supergravity ultraviolet finite?,}
  Phys.\ Lett.\  B {\bf 644} (2007) 265,
  {\tt hep-th/0611086}.


\bibitem{bsgf2}
  Z.~Bern, J.~J.~Carrasco, L.~J.~Dixon, H.~Johansson, D.~A.~Kosower and R.~Roiban,
 {\it Three-Loop Superfiniteness of ${\cal N}=8$ Supergravity,}
  Phys.\ Rev.\ Lett.\  {\bf 98} (2007) 161303,
  {\tt hep-th/0702112}.






\bibitem{bsgf3}
  Z.~Bern, J.~J.~Carrasco, D.~Forde, H.~Ita and H.~Johansson,
  {\it Unexpected Cancellations in Gravity Theories,}
  Phys.\ Rev.\  D {\bf 77} (2008) 025010,
 {\tt 0707.1035 [hep-th]}.




\bibitem{Bern:2009kd}
  Z.~Bern, J.~J.~Carrasco, L.~J.~Dixon, H.~Johansson and R.~Roiban,
  {\it The Ultraviolet Behavior of ${\cal N}=8$ Supergravity at Four Loops,}
  {\tt 0905.2326 [hep-th]}.




\bibitem{bddk} 
Z.~Bern, L.~J.~Dixon, D.~C.~Dunbar and D.~A.~Kosower, {\it One-Loop N-Point Gauge Theory Amplitudes, Unitarity And Collinear Limits,}
    Nucl.\ Phys.\ B {\bf 425}, 217 (1994), {\tt hep-ph/9403226}.



\bibitem{Bern:1998sv}
  Z.~Bern, L.~J.~Dixon, M.~Perelstein and J.~S.~Rozowsky,
  {\it Multi-leg one-loop gravity amplitudes from gauge theory,}
  Nucl.\ Phys.\  B {\bf 546} (1999) 423,
  {\tt hep-th/9811140}.



\bibitem{BjerrumBohr:2006yw}
  N.~E.~J.~Bjerrum-Bohr, D.~C.~Dunbar, H.~Ita, W.~B.~Perkins and K.~Risager,
 {\it The no-triangle hypothesis for ${\cal N}=8$ supergravity,}
  JHEP {\bf 0612} (2006) 072,
  {\tt hep-th/0610043}.

%
\bibitem{BjerrumBohr:2008vc}
  N.~E.~J.~Bjerrum-Bohr and P.~Vanhove,
  {\it Explicit Cancellation of Triangles in One-loop Gravity Amplitudes,}
  JHEP {\bf 0804} (2008) 065,
  {\tt 0802.0868 [hep-th]}.



\bibitem{cahk}
  N.~Arkani-Hamed, F.~Cachazo and J.~Kaplan,
  {\it What is the Simplest Quantum Field Theory?,}
  {\tt 0808.1446 [hep-th]}.


\bibitem{Badger:2008rn}
  S.~Badger, N.~E.~J.~Bjerrum-Bohr and P.~Vanhove,
  {\it Simplicity in the Structure of QED and Gravity Amplitudes,}
  JHEP {\bf 0902} (2009) 038,
  {\tt 0811.3405 [hep-th]}.




\bibitem{ben}
  P.~Benincasa, C.~Boucher-Veronneau and F.~Cachazo,
 {\it Taming tree amplitudes in general relativity,}
  JHEP {\bf 0711} (2007) 057,
{\tt hep-th/0702032}.

\bibitem{ah1}
  N.~Arkani-Hamed and J.~Kaplan,
 {\it On Tree Amplitudes in Gauge Theory and Gravity,}
  JHEP {\bf 0804} (2008) 076,
{\tt 0801.2385 [hep-th]}.



\bibitem{Binoth:2002xg}
  T.~Binoth, E.~W.~N.~Glover, P.~Marquard and J.~J.~van der Bij,
  {\it Two-loop Corrections to Light-by-Light Scattering in Supersymmetric QED},
  JHEP {\bf 0205}, 060 (2002),  {\tt hep-ph/0202266}.





\bibitem{Naculich:2008ew}
  S.~G.~Naculich, H.~Nastase and H.~J.~Schnitzer,
  {\it Two-loop graviton scattering relation and IR behavior in ${\cal N}=8$
  supergravity,}
  Nucl.\ Phys.\  B {\bf 805} (2008) 40, 
  {\tt 0805.2347 [hep-th]}.

\bibitem{Brandhuber:2008tf}
  A.~Brandhuber, P.~Heslop, A.~Nasti, B.~Spence and G.~Travaglini,
  {\it Four-point Amplitudes in ${\cal N}=8$ Supergravity and Wilson Loops,}
  Nucl.\ Phys.\  B {\bf 807} (2009) 290, 
  {\tt 0805.2763 [hep-th]}.



\bibitem{klov}
  A.~V.~Kotikov, L.~N.~Lipatov, A.~I.~Onishchenko and V.~N.~Velizhanin,
  {\it Three-loop universal anomalous dimension of the Wilson operators in ${\cal N}=4$
  SUSY Yang-Mills model,}
  Phys.\ Lett.\  B {\bf 595} (2004) 521
  [Erratum-ibid.\  B {\bf 632} (2006) 754], 
  {\tt hep-th/0404092}.

\bibitem{kinoshitacvitanovic}
  P.~Cvitanovi\'{c} and T.~Kinoshita,
  {\it Sixth Order Magnetic Moment Of The Electron,}
  Phys.\ Rev.\  D {\bf 10}, 4007 (1974).


\bibitem{cvitanovic}P.~Cvitanovi\'{c}, {\it Acceptance speech, 1993 NKH Research Prize in Physics, Dansk Fysisk Selskab
{\AA}rsm{\o}de.
}





\bibitem{Karplus:1950zz}
  R.~Karplus and M.~Neuman,
  {\it The scattering of light by light},
  Phys.\ Rev.\  {\bf 83}, 776 (1951).

\bibitem{Karplus:1950zza}
  R.~Karplus and M.~Neuman,
  {\it Non-Linear Interactions Between Electromagnetic Fields},
  Phys.\ Rev.\  {\bf 80}, 380 (1950).


\bibitem{Bern:2001dg}
  Z.~Bern, A.~De Freitas, L.~J.~Dixon, A.~Ghinculov and H.~L.~Wong,
{\it QCD and QED corrections to light-by-light scattering},
  JHEP {\bf 0111}, 031 (2001), {\tt hep-ph/0109079}.



\bibitem{fusing}
  Z.~Bern, L.~J.~Dixon, D.~C.~Dunbar and D.~A.~Kosower,
{\it Fusing gauge theory tree amplitudes into loop amplitudes,}
  Nucl.\ Phys.\  B {\bf 435}, 59 (1995),
  {\tt hep-ph/9409265}.





\bibitem{Mahlon:1993fe}
  G.~Mahlon,
  {\it One loop multi - photon helicity amplitudes}, Phys.\ Rev.\  D {\bf 49}, 2197 (1994),
  {\tt hep-ph/9311213}.


\bibitem{Mahlon2}
G.~Mahlon,
  {\it Multi - Photon Production At High-Energies In The Standard Model. 2,}
  Phys.\ Rev.\  D {\bf 47}, 1812 (1993), 
  {\tt hep-ph/9210214}.

\bibitem{Binoth:2007ca}
  T.~Binoth, G.~Heinrich, T.~Gehrmann and P.~Mastrolia,
  {\it Six-Photon Amplitudes},
  Phys.\ Lett.\  B {\bf 649}, 422 (2007),  {\tt hep-ph/0703311}.

\bibitem{Bernicot:2007hs}
  C.~Bernicot and J.~P.~Guillet,
  {\it Six-Photon Amplitudes in Scalar QED},
  JHEP {\bf 0801}, 059 (2008),  {\tt 0711.4713 [hep-ph]}.


\bibitem{csw}
F.~Cachazo, P.~Svr\v{c}ek and E.~Witten,
{\it MHV vertices and tree amplitudes in gauge theory},
JHEP {\bf 0409} (2004) 006, {\tt hep-th/0403047}.


\bibitem{ozeren}
K.~J.~Ozeren and W.~J.~Stirling,
{\it MHV techniques for QED processes},
JHEP {\bf 0511} (2005) 016, {\tt hep-th/0509063}.


\bibitem{bst} 
A.~Brandhuber, B.~Spence and G.~Travaglini,
{\it One-Loop Gauge Theory Amplitudes in \mbox{${\cal N}=4$} super Yang-Mills from MHV Vertices}, Nucl.\ Phys.\ B {\bf 706}, 150 (2005), {\tt  hep-th/0407214}. 


\bibitem{Bedford:2004py}
J.~Bedford, A.~Brandhuber, B.~J.~Spence and G.~Travaglini,
{\it A twistor approach to one-loop amplitudes in ${\cal N}=1$ supersymmetric
Yang-Mills theory},  Nucl.\ Phys.\  B {\bf 706}, 100 (2005),
{\tt hep-th/0410280}.



\bibitem{Bedford:2004nh}
J.~Bedford, A.~Brandhuber, B.~J.~Spence and G.~Travaglini,
{\it Non-supersymmetric loop amplitudes and MHV vertices},  Nucl.\ Phys.\  B {\bf 712}, 59 (2005),
{\tt hep-th/0412108}.





\bibitem{Bern:2006vw}
  Z.~Bern, M.~Czakon, D.~A.~Kosower, R.~Roiban and V.~A.~Smirnov, 
  {\it Two-loop iteration of five-point ${\cal N}=4$ super-Yang-Mills amplitudes},  
Phys.\ Rev.\ Lett.\  {\bf 97}, 181601 (2006),  {\tt hep-th/0604074}.
  
\bibitem{Cachazo:2006tj}
  F.~Cachazo, M.~Spradlin and A.~Volovich, 
  {\it Iterative structure within the five-particle two-loop amplitude},  Phys.\ Rev.\  D {\bf 74}, 
045020 (2006), 
{\tt hep-th/0602228}.
 
  
  

\bibitem{Alday:2007he}
  L.~F.~Alday and J.~Maldacena,
 {\it Comments on gluon scattering amplitudes via AdS/CFT,}
  JHEP {\bf 0711} (2007) 068, 
  {\tt 0710.1060 [hep-th]}.
 
\bibitem{seven}
  Z.~Bern, L.~J.~Dixon, D.~A.~Kosower, R.~Roiban, M.~Spradlin, C.~Vergu and A.~Volovich,
{\it The Two-Loop Six-Gluon MHV Amplitude in Maximally Supersymmetric Yang-Mills
  Theory,}
  Phys.\ Rev.\  D {\bf 78}, 045007 (2008),
  {\tt 0803.1465 [hep-th]}.

\bibitem{dhksbum}
J.~M.~Drummond, J.~Henn, G.~P.~Korchemsky and E.~Sokatchev,
{\it The hexagon Wilson loop and the BDS ansatz for the six-gluon amplitude,}
Phys.\ Lett.\  B {\bf 662}, 456 (2008),  {\tt 0712.4138 [hep-th]}.

\bibitem{dhks6}
J.~M.~Drummond, J.~Henn, G.~P.~Korchemsky and E.~Sokatchev,
{\it Hexagon Wilson loop = six-gluon MHV amplitude,}
{\tt 0803.1466 [hep-th].}



\bibitem{Mangano:1990by}
M.~L.~Mangano and S.~J.~Parke,
{\it Multi-Parton Amplitudes in Gauge Theories},
Phys.\ Rept.\  {\bf 200}, 301 (1991), {\tt hep-th/0509223}.




\bibitem{bern1}
Z.~Bern, G.~Chalmers, L.~J.~Dixon and D.~A.~Kosower,
{\it One loop N gluon amplitudes with maximal helicity violation via collinear limits},
Phys.\ Rev.\ Lett.\  {\bf 72}, 2134 (1994),
{\tt hep-ph/9312333}.

\bibitem{bern2}
Z.~Bern, L.~J.~Dixon and D.~A.~Kosower,
{\it One Loop Corrections To Two Quark Three Gluon Amplitudes},
Nucl.\ Phys.\  B {\bf 437}, 259 (1995),
{\tt hep-ph/9409393}.

\bibitem{Bern:1993kr}
Z.~Bern, L.~J.~Dixon and D.~A.~Kosower,
{\it Dimensionally regulated pentagon integrals,}
Nucl.\ Phys.\  B {\bf 412} (1994) 751, {\tt hep-ph/9306240}.




\bibitem{Bern:1993mq}
  Z.~Bern, L.~J.~Dixon and D.~A.~Kosower,
  {\it One loop corrections to five gluon amplitudes,}
  Phys.\ Rev.\ Lett.\  {\bf 70} (1993) 2677, 
  {\tt hep-ph/9302280}.


\bibitem{Brandhuber:2008cy}
A.~Brandhuber and M.~Vincon,
{\it MHV One-Loop Amplitudes in Yang-Mills from Generalised Unitarity},
JHEP {\bf 0811}, 078 (2008), {\tt 0805.3310 [hep-ph]}.

\bibitem{Bern:1997sc}
  Z.~Bern, L.~J.~Dixon and D.~A.~Kosower,
  {\it One-loop amplitudes for $e^{+}e^{-}$ to four partons,}
  Nucl.\ Phys.\  B {\bf 513} (1998) 3,
  {\tt hep-ph/9708239}.

\bibitem{bcfgen}
R.~Britto, F.~Cachazo and B.~Feng,
{\it Generalized unitarity and one-loop amplitudes in ${\cal N}=4$  super-Yang-Mills,}
Nucl.\ Phys.\  B {\bf 725}, 275 (2005), {\tt hep-th/0412103}.





\bibitem{ir6}
  L.~Magnea and G.~Sterman,
  {\it Analytic continuation of the Sudakov form-factor in QCD,}
  Phys.\ Rev.\  D {\bf 42} (1990) 4222.





\bibitem{am}
  L.~F.~Alday and J.~M.~Maldacena, {\it Gluon scattering amplitudes at strong coupling},  JHEP {\bf 0706}, 064 (2007), {\tt 0705.0303 [hep-th]}.

\bibitem{dks1}
  J.~M.~Drummond, G.~P.~Korchemsky and E.~Sokatchev, {\it Conformal properties of four-gluon planar amplitudes and Wilson loops},   
Nucl.\ Phys.\  B {\bf 795}, 385 (2008), {\tt 0707.0243 [hep-th]}.


\bibitem{bht}
  A.~Brandhuber, P.~Heslop and G.~Travaglini,
  {\it MHV Amplitudes in ${\cal N}=4$ Super Yang-Mills and Wilson Loops,}
  Nucl.\ Phys.\  B {\bf 794},  231 (2008),
  {\tt 0707.1153 [hep-th]}.



\end{thebibliography}
\end{document}